\documentclass[conference,usenatbib]{basi}
\usepackage[T1]{fontenc}
\usepackage[british]{babel}
\usepackage[varg]{txfonts}
\usepackage{rotating}
\usepackage{dcolumn}

\begin{document}
\title[Core-Collapse SNe]{Core-Collapse supernovae and its progenitors}
\author[Subhash Bose]%
       {Subhash Bose\thanks{email: \texttt{bose@aries.res.in}},
       Brijesh Kumar and Kuntal Misra\\
       Aryabhatta Research Institute of observation sciencES, Manora Peak, Nainital 263002, India}

\pubyear{2015}
\volume{12}
\pagerange{\pageref{firstpage}--\pageref{lastpage}}

\date{Received --- ; accepted ---}

\maketitle
\label{firstpage}

\begin{abstract}
Massive stars unable to sustain gravitational collapse, at the end of nuclear burning stage, turns out into core-collapse supernovae, leaving behind compact objects like neutron stars or black holes.  The progenitor properties like mass and metallicity primarily governs the explosion parameters and type of compact remnant.  In this contribution we summarize observational study of three Core Collapse type IIP SNe 2012aw, 2013ab and 2013ej, which are rigorously observed from ARIES and other Indian observatories and discuss their progenitor and explosion properties.
\end{abstract}

\vspace{-1em}


\section{Introduction}\label{s:intro}
\vspace{-0.5em}

Evolution of massive stars leads into core-collapse supernovae leaving behind compact objects like
neutron stars and black holes. The properties of progenitor star (primarily the  mass and metallicity) govern the 
parameters of explosion and the type of compact remnant. Type II core-collapse supernovae originate from 
stars with zero-age-main-sequence (ZAMS) mass $ \rm 8<M<25 \rm M\odot$ \citep{2003ApJ...591..288H,2004MNRAS.353...87E}
and retains sufficient hydrogen at the time of explosion which is prominently visible in their spectra.

\section{Summary of three CCSNe observed from ARIES}
\begin{figure}
\centering
\includegraphics[width=0.45\linewidth]{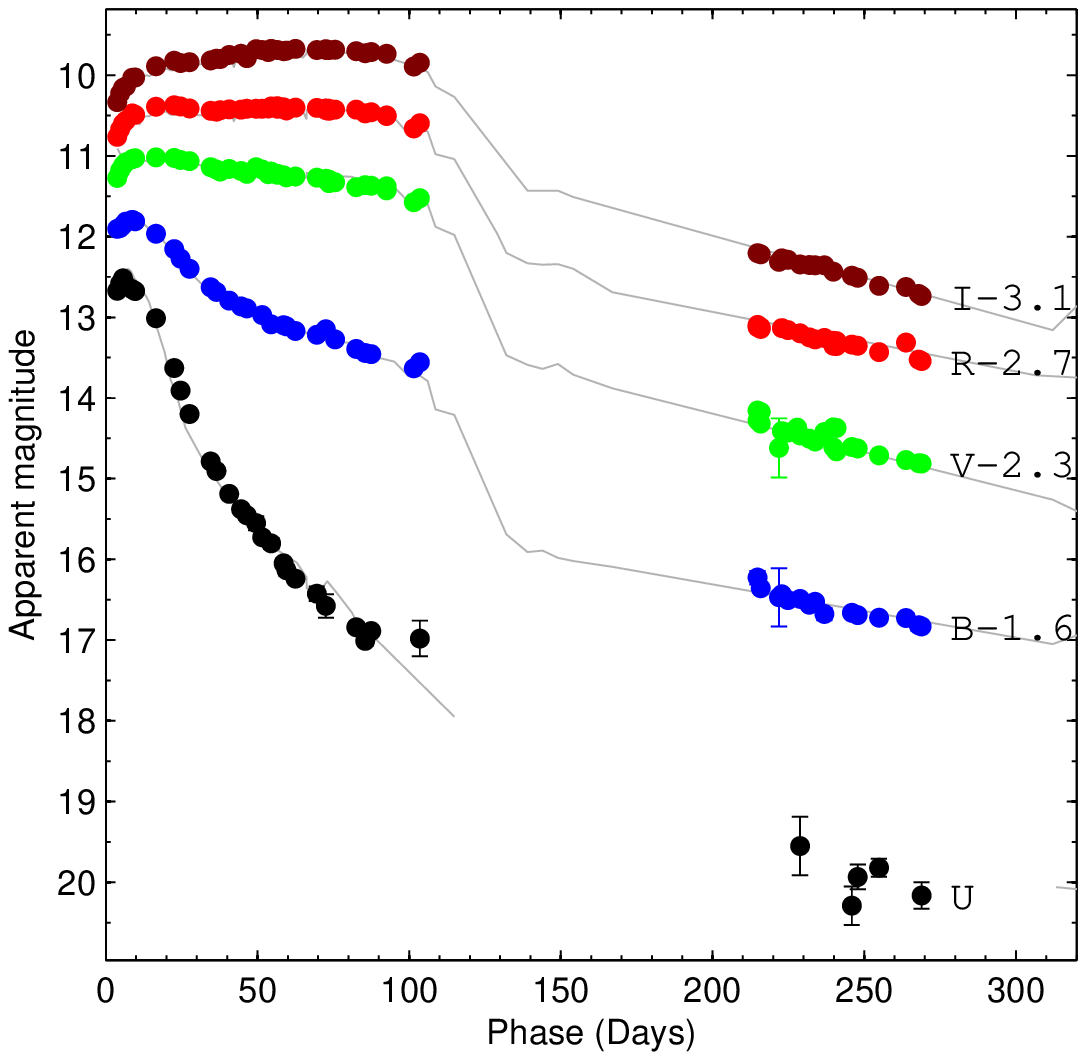}
\includegraphics[width=0.45\linewidth]{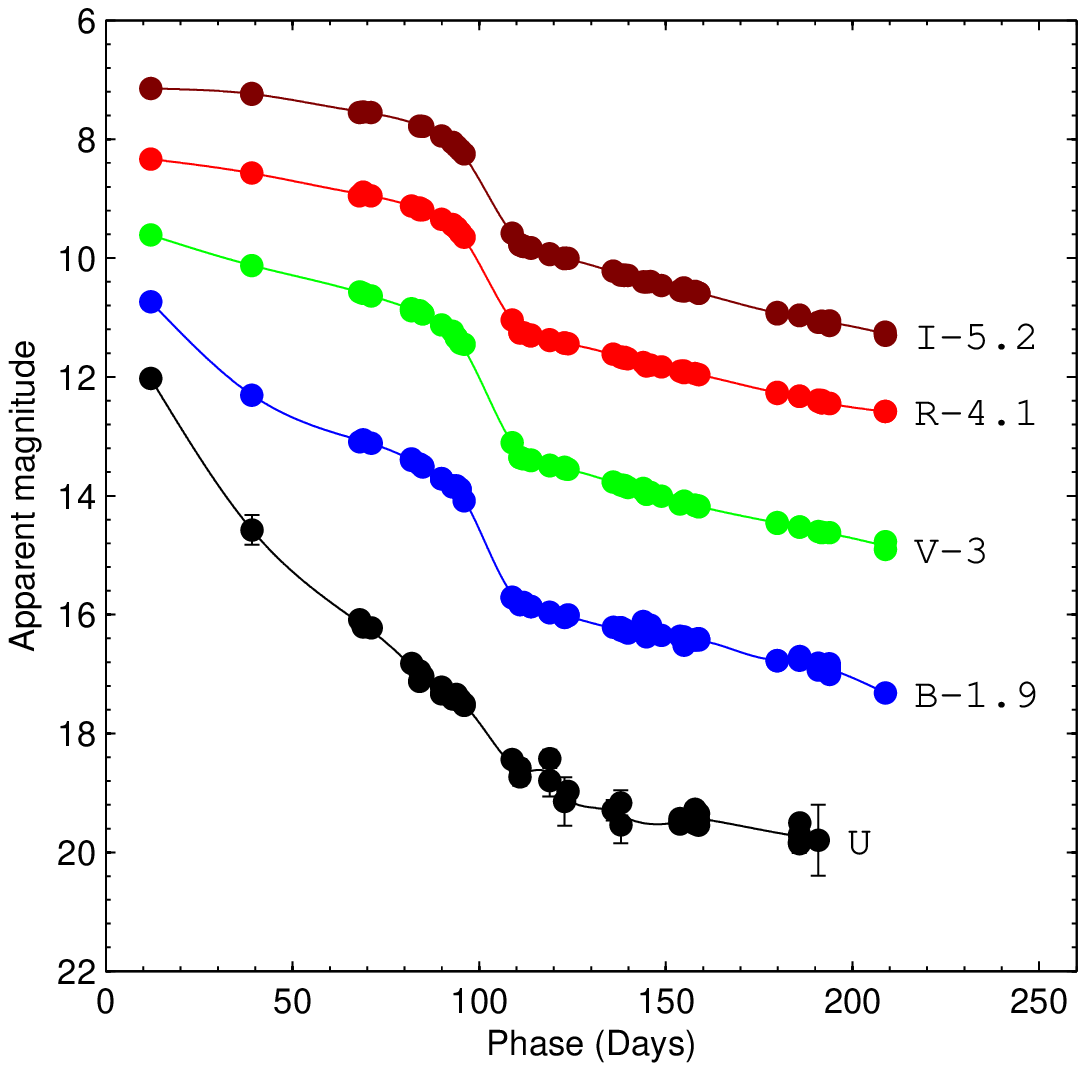}
\caption{Light curves of SN 2012aw \citep[left;][]{2013MNRAS.433.1871B} and SN 2013ej \citep[right;][]{2015ApJ...806..160B} in \textit{UBVRI} bands are shown.}
\label{figure1}
\end{figure}

\vspace{-0.5em}
SNe 2012aw, 2013ab and 2013ej are thee bright CCSNe which has been rigorously observed, both photometrically and spectroscopically, from ARIES telescopes and other observatories across India including Himalayan Chandra telescope and IUCAA Girawali observatory. The light curves of SNe 2012aw and 2013ej are shown in Fig.~\ref{figure1}. SNe 2012aw and 2013ab belongs to normal class of type IIP SNe, whereas SN 2013ej shows a steeper plateau phase resembling type IIL or intermediate slope SNe. Spectra of SNe 2012aw and 2013ej shows high velocity features in H~\textsc{i} P-Cygni profiles from mid-plateau plateau onwards, which is a possible signature of ejecta-CSM interaction.
The observations for these supernova has been used to perform hydrodynamical or semi-analytical modelling to estimate progenitor properties and explosion energy, the results are listed in Table~\ref{table}. 
Hydrodynamical model yield more accurate result, while semi-analytical model has advantage of being less computational resource intensive. However, results from both models are consistent within the limit of errors.
\vspace{-1.0em}
\begin{table}[!h]
\centering
\fontsize{9pt}{3.5mm}\selectfont
\caption{Progenitor properties and explosion energy estimated from modelling}
\label{table}
\begin{tabular}{cccccc}
\hline
 Supernova & Mass  & Radius & Energy & Reference & Modelling \\ 
  & $ M\odot $ & $ R\odot $  & $ 10^{51} $ergs &   & type\\ \hline
 SN 2012aw & 22 & 430 & 1.5 & \cite{2014ApJ...787..139D} & Hydrodynamical \\ 
 SN 2013ab & 9 & 600 & 0.35 & \cite{2015MNRAS.450.2373B} & Hydrodynamical \\ 
 SN 3013ej & 14 & 450 & 2.10 & \cite{2015ApJ...806..160B} & Semi-analytical \\ 
\hline 
\end{tabular} 
\end{table}
\vspace{-1.5em} 

The estimated progenitor properties indicate that they all originate from red supergiants and stellar evolutionary models \citep[e,g.,][]{2003ApJ...591..288H} suggests that stars of these mass range would leave behind neutron star as compact remnant. 
\vspace{-0.2em} 


\begin{thebibliography}{}
%
%
%
%
%
%
%
%


 \bibitem[Bose et al.(2013)Bose et al.]{2013MNRAS.433.1871B} 
 Bose S., et al., 2013, MNRAS, 433, 1871 
 
 \bibitem[Bose et al.(2015a)Bose et al.]{2015MNRAS.450.2373B} 
 Bose S., et al., 2015a, MNRAS, 450, 2373 
 
 \bibitem[Bose et al.(2015b)Bose et al.]{2015ApJ...806..160B} 
 Bose S., et al., 2015b, ApJ, 806, 160 
 
  \bibitem[Dall'Ora et al.(2014)Dall'Ora et al.]{2014ApJ...787..139D} 
  Dall'Ora M., et al., 2014, ApJ, 787, 139 
  
 
  \bibitem[Eldridge \& Tout(2004)Eldridge \& Tout]{2004MNRAS.353...87E} 
  Eldridge J.~J., Tout C.~A., 2004, MNRAS, 353, 87 
 
\bibitem[Heger et al.(2003)Heger et al.]{2003ApJ...591..288H} 
 Heger A., Fryer C.~L., Woosley S.~E., Langer N., Hartmann D.~H., 2003, ApJ, 591, 288 



%
%
%
%

\end{thebibliography}
\end{document}